\documentclass[aps,prl,twocolumn,superscriptaddress,showpacs,showkeys,30final]{revtex4-2}
\usepackage{amsmath,latexsym,amsfonts,amssymb}
\usepackage{natbib}
\usepackage{epsfig}
\usepackage{graphicx}
\usepackage{inputenc}
\usepackage{color}
\usepackage{setspace}
\usepackage{multirow}
\usepackage{array}
\usepackage{tabularx}
\usepackage{bm}
\usepackage{color}
\usepackage{xcolor}
\usepackage{gensymb}
\usepackage{xcolor}
\usepackage[left]{lineno}
\usepackage{float} 
\usepackage{ulem}
\usepackage{siunitx}

\newcommand\st{\bgroup\markoverwith
  {\textcolor{blue}{\rule[.35ex]{5pt}{1.1pt}}}\ULon}

\newcommand{\bfse}{BaFe$_2$Se$_3$}	

\begin{document}

\title{\bfse\: a quasi-unidimensional non-centrosymmetric superconductor}

\author{S. Deng$^{*}$}
\affiliation{$^{*}$These authors contributed equally to this work.}
\affiliation{Institut Laue Langevin, 38000 Grenoble, France}

\author{A. Roll$^{*}$}
\affiliation{Universit\'e Paris-Saclay, CNRS, Laboratoire de Physique des Solides, 91405, Orsay, France.}
\affiliation{Universit\'e Paris-Saclay, CNRS-CEA, Laboratoire L\'eon Brillouin, 91191, Gif sur Yvette, France}

\author{W. G. Zheng$^{*}$}
\affiliation{Universit\'e Paris-Saclay, CNRS, Laboratoire de Physique des Solides, 91405, Orsay, France.}

\author{G. Giri}
\affiliation{Institut Laue Langevin, 38000 Grenoble, France}
\affiliation{Current address: Physics Dept., VIT Bhopal University, Bhopal, India}

\author{T. Vasina}
\affiliation{Institut Laue Langevin, 38000 Grenoble, France}

\author{D. Bounoua}
\affiliation{Universit\'e Paris-Saclay, CNRS-CEA, Laboratoire L\'eon Brillouin, 91191, Gif sur Yvette, France}

\author{P. Fertey}
\affiliation{Synchrotron SOLEIL, L\'\ Orme des Merisiers, Saint Aubin BP 48, 91192, Gif-sur-Yvette, France}

\author {M. Verseils}
\affiliation{Synchrotron SOLEIL, L\'\ Orme des Merisiers, Saint Aubin BP 48, 91192, Gif-sur-Yvette, France}

\author {C. Bellin}
\affiliation{Universit\'e Pierre et Marie Curie, IMPMC, CNRS UMR7590, 4 Place Jussieu, 75005 Paris, France}

\author{A. Forget}
\affiliation{Universit\'e Paris-Saclay, CEA, CNRS, SPEC, 91191, Gif-sur-Yvette, France.}

\author{D. Colson}
\affiliation{Universit\'e Paris-Saclay, CEA, CNRS, SPEC, 91191, Gif-sur-Yvette, France.}

\author{P. Foury-Leylekian}
\affiliation{Universit\'e Paris-Saclay, CNRS, Laboratoire de Physique des Solides, 91405, Orsay, France.}

\author {M.B. Lepetit}
\affiliation{Institut N\'eel, CNRS, 38042 Grenoble, France}
\affiliation{Institut Laue Langevin, 38000 Grenoble, France}

\author{V. Bal\'edent}
\affiliation{Universit\'e Paris-Saclay, CNRS, Laboratoire de Physique des Solides, 91405, Orsay, France.}
\affiliation{Institut universitaire de France (IUF)}
\email [Corresponding author: ] {victor.baledent@universite-paris-saclay.fr}

\date{\today}

\begin{abstract}
  The spin-ladder compounds of the BaFe$_2$X$_3$ (X = chalcogen) family may be viewed as dimensional reductions—along stripe-like motifs—of the two-dimensional iron-based pnictide planes extensively studied since 2006. Remarkably, despite their reduced dimensionality, these materials retain the capacity for unconventional ground states, exemplified by the emergence of superconductivity in \bfse\ under applied pressure beyond 10 GPa, following a structural phase transition at 4 GPa. Here, we report a comprehensive investigation combining high-resolution single-crystal X-ray diffraction, infrared spectroscopy, and ab initio calculations, which together elucidate the true crystallographic nature of this pressure-induced superconducting phase. While X-ray diffraction alone reveals a symmetry lowering from the widely accepted orthorhombic $Cmcm$ group to a monoclinic structure, it lacks sufficient sensitivity to resolve the precise space group. By integrating vibrational spectroscopy with density functional theory, we provide unambiguous evidence that the high-pressure phase is non-centrosymmetric, adopting the polar space group $P2_1$. These findings not only revise the structural assignment of \bfse\ in its superconducting state but also establish its non-centrosymmetric character—an essential ingredient for potential unconventional pairing mechanisms—thereby opening new perspectives on the interplay between lattice symmetry, dimensionality, and superconductivity in iron-based materials.  
\end{abstract}

\maketitle
Unconventional superconductivity remains one of the most emblematic quantum properties in solid-state physics, resisting any consensus on a phenomenological or microscopic model \cite{Scalapino2012, Keimer2015, Stewart2017}. Initiated by the discovery of high-critical temperature cuprates, this pursuit of elevated temperatures has led to the exploration of new materials such as iron-based pnictides, heavy fermion compounds, and more recently, nickelates ~\cite{Stewart2011}. While the mechanism behind Cooper pair formation in these families remains elusive, there is a broad consensus that understanding superconductivity requires investigating phases and orders neighboring the superconducting dome ~\cite{Fernandes2014, Fradkin2015}. Indeed, to elucidate pairing mechanisms or test proposed models, it is pertinent to observe the presence of competing or intertwined order parameters.
This has been done in particular in numerous quasi-two-dimensional (2D) iron-based superconductors such as FeSe or BaFe$_2$As$_2$, whose square-lattice planes and layered architectures have been extensively studied both structurally and magnetically \cite{Johnston2010,Stewart2011,Kordyuk2012, Baledent2012, Baledent2015}. These studies have revealed how structural and magnetic degrees of freedom, often intimately linked, evolve upon approaching the superconducting dome via chemical doping or applied pressure.
Recently, new spin-ladder iron-based systems have demonstrated superconductivity under pressure. With a general formula of BaFe$_2$X$_3$ (X=S and Se), they represent the first quasi-one-dimensional iron-based superconductors devoid of a square-lattice motif~\cite{Takahashi2015,Ying2017}. The quasi-one-dimensional character of these compounds originates from their two-leg ladder crystal structure and is reflected in strongly anisotropic transport properties and low-dimensional magnetic correlations~\cite{Lei2011,Nambu2012,Mourigal2015,Ying2017}. In this context, these two compounds offer a unique opportunity to investigate how reducing dimensionality from 2D to 1D affects the delicate balance between magnetism, structure, and superconductivity.

Consisting of two iron ladders per unit cell, the \bfse\ and BaFe$_2$S$_3$ compounds were respectively reported to crystallize in the $Pnma$ and $Cmcm$ space groups~\cite{hong1972crystal}. In the $Cmcm$ structure, the ladders, formed by edge-sharing FeX$_4$ tetrahedra along the $c$ axis, are perfectly contained in the ($a$,$c$) plane, whereas a tilt is observed in $Pnma$. Interestingly, \bfse\ is reported to undergo a transition from $Pnma$ to $Cmcm$ above 3--4~GPa\cite{Svitlyk2019}, suggesting a common atomic structure between the two compounds close to the superconducting phase. This apparent universality of the neighboring phases of the superconducting dome is reinforced by the presence of the same magnetic order, as measured by neutron diffraction under pressure~\cite{Zheng2022}.
However, a more thorough investigation on single crystals of \bfse, utilizing X-ray diffraction, infrared spectroscopy, and {\it ab-initio} calculations, revealed that the actual space group at ambient pressure is $Pm$, a significantly lower symmetry than the previously reported average $Pnma$ space group~\cite{Zheng2020,Weseloh2022}. Understanding the symmetries at play in these compounds is therefore crucial for comprehending their physical properties, especially since such symmetry breaking can endow \bfse\ with multiferroic behavior. An even more intriguing possibility is that the loss of inversion symmetry persists into the superconducting phase, placing \bfse\ within the rare class of non-centrosymmetric superconductors~\cite{GorkovRashba2001,Samokhin2009,Smidman2017}.

Until now, very limited information has been available regarding the crystallographic structure of the superconducting phase itself, despite its crucial importance for understanding the nature of superconductivity in this system. In light of the ambient-pressure results, it is therefore essential to investigate the structure of \bfse\ in single crystals under combined high-pressure and low-temperature conditions. In this Letter, we present a study of \bfse\ under pressure, combining X-ray diffraction (XRD), infrared (IR) and Raman spectroscopies, and {\it ab-initio} calculations. Our results reveal that the high-pressure space group of \bfse\ in the phase where superconductivity emerges is $P2_1$. This symmetry establishes \bfse\ as one of the very few known examples of a quasi-one-dimensional non-centrosymmetric superconductor.

\paragraph{Experimental methods.}
Single crystals of \bfse\ were grown using the self-flux method as described in Ref.~\cite{Zheng2020}. We emphasize that the samples used in this study originate from the same batch that was extensively characterized in previous works. Compositional uniformity was verified by energy-dispersive X-ray spectroscopy (EDX), showing a nominal stoichiometry within 2\% uncertainty (see SI of Ref.~\cite{Zheng2020}), and scanning electron microscopy (SEM) confirmed the absence of structural defects at the microscale~\cite{Zheng2023}. Although structural symmetry, particularly non-centrosymmetry, can be sensitive to stoichiometry and sample quality, no deviation from the reported superconducting composition or structure was detected. Despite the absence of low-temperature resistivity measurements under pressure showing superconductivity, we are confident that the observed non-centrosymmetric structure corresponds to a superconducting phase representative of this material family. For each experiment, one or several single crystals of \bfse\ were loaded in a membrane-driven diamond anvil cell (DAC), using a stainless steel gasket with pressure transmitting media adapted to each technique (see details in the Supplementary Information (SI) \cite{SI}). The pressure inside the diamond anvil cell was measured using the ruby fluorescence technique, yielding a 5\% relative accuracy on the pressure value \cite{Shen2020}. XRD and IR reflectivity were performed on the CRISTAL and AILES beamline respectively, at synchrotron SOLEIL. 
Due to technical constraints associated with the cryostat used for measurements in diamond anvil cells required for high-pressure experiments, the lowest temperatures achieved were not identical across techniques: 10~K for diffraction measurements and 50~K for infrared spectroscopy. However, as shown in our previous work at ambient pressure \cite{Weseloh2022}, phonon energies exhibit minimal variation between 50 K and 10 K. 

Raman scattering experiments were performed at IMPMC (Paris) using a 534~nm laser source with a fixed power of 80~mW in a backscattering geometry. Measurements were carried out at 300~K, as the Raman signal is particularly weak in this family of compounds already at ambiant conditions, consistent with previous reports~\cite{Popovic2015}, preventing us to measure at both low temperature and high pressure despite several attempts. Polarized Raman spectra were successfully measured under pressure using well-defined scattering geometries, expressed in Porto notation as $(a,a)$ and $(c,c)$ configurations, where both incident and scattered electric fields are aligned along the crystallographic $\mathbf a$ and $\mathbf c$ directions, respectively (in $Cmcm$ notation).

The {\it ab-initio} calculations were done using Density Functional Theory (DFT). In order to best account for the eletronic correlation at the Fermi level, we used the hybrid B3LYP functional~\cite{B3LYP} and a localized basis set.  An atomic gaussian basis of $3\zeta+P$ quality was used for the Fe and Se atoms~\cite{pob-TZVP-rev2}. The Ba atoms were represented using a relativistic core pseudo-potential of the Stuttgard group~\cite{PseudoR13} and the associated basis set where the diffuse functions exponents were taken as 1.2 as recommended by Scuseria {\it et al}~\cite{HSE}. The calculations were done using the CRYSTAL23 code~\cite{CRYSTAL23}.  The shrinking factor was set to $8\times 12 \times 4$ resulting in a sampling net of 126 $\bf k$-points. The stripe spin ordering found in neutron scattering experiments of \bfse\ under pressure~\cite{Zheng2022} was used in all calculations.

\paragraph{X-ray diffraction.} 
The most comprehensive study of structural properties in this compound is provided in Ref.~\onlinecite{Svitlyk2019}, where room temperature X-Ray data under pressure up to 50~GPa are provided. Between 0~--~3.5\;GPa, V. Svitlyk {\it et al} propose a $Pnma$ space group, between 3.5~--~16.6\;GPa a $Cmcm$ group and above 16.6\;GPa an isostructural $Cmcm$ phase where the {\bf b} axis collapses while the {\bf a} and {\bf c} ones expand \cite{Svitlyk2019}. 

One should however remember that in these compounds diffraction measurements often lead to average structures, rather than true space groups. Indeed, while at ambient pressure  X-Ray were suggesting a $Pnma$ group~\cite{hong1972crystal,Svitlyk2019}, our high resolution X-Ray and lattice dynamic investigations clearly identified small symmetry breakings and the true space group to be $Pm$. 

We collected thousands of Bragg reflections at selected pressure and temperature values; see details in SI~\cite{SI}. 

\begin{figure}[h!]
\includegraphics[width=0.999\linewidth, angle=0]{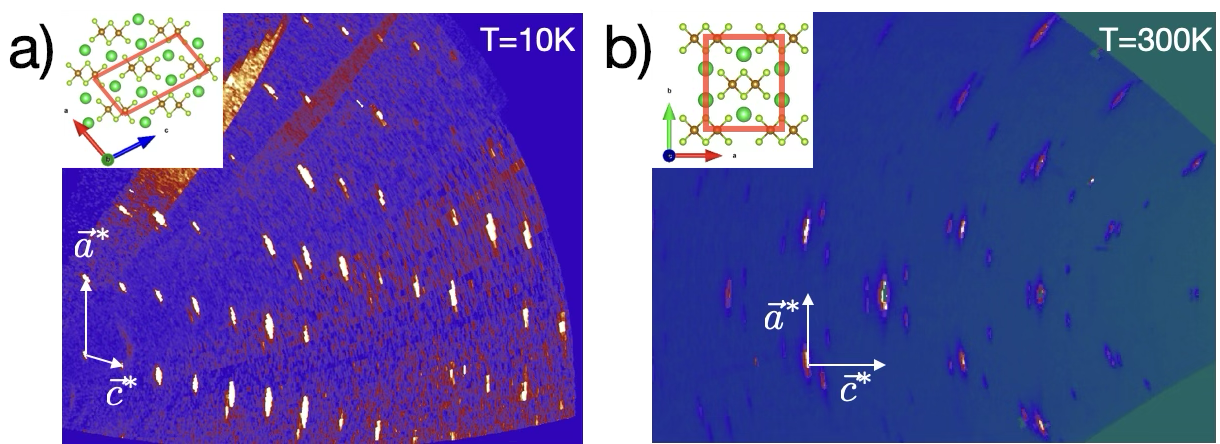}
\caption{(Color online) Reconstructions of the reciprocal space at 12 GPa (a) at low temperature in the (H,0,L) plane perpendicular to the ladders in the monoclinic setting and (b) at ambiant temperature in the (H,K,0) plane perpendicular to the ladders in the $Cmcm$ setting. Insets show the lattice unit in red with the direct lattice vectors in their respective settings.}
\label{reconstruction}
\end{figure}

At 300\;K, above the 3-4\;GPa transition and up to 12\;GPa, our X-Ray data  confirm the reported~\cite{Svitlyk2019} $Cmcm$ space group, as we observed no intensity on forbidden reflections (see Fig.~\ref{reconstruction}b).

At 10\;K, the data collected at 5 and 12\;GPa indicate that the space group cannot be   $Cmcm$. Specifically, we observed intensities for i) $H+K$ odd, even for $L=0$, indicating a break in $C$ centering and of the $\{m_{00z}, \frac{1}{2} \frac{1}{2}  \frac{1}{2}\}$ glide mirror and ii) $(H,0,L)$ with $L$ odd indicating a breaking of the $\mathbf c$ glide mirror perpendicular to the $\mathbf b$ axis iii) no intensity on the  $(0,0,L)$ with L odd peaks and iv) a clear monoclinic structure with a significant angle $\approx 103^\circ$ and the unique axis along  the $\mathbf{c}$ axis of the $Cmcm$. As a consequence the space group  for \bfse\ at high-pressure and low temperature should be either $P2_1/m$ or $P2_1$ with the unique axis along  the $\mathbf{c}$ axis of the $Cmcm$ (see SI~\cite{SI} for a detailled symmetry analysis).

We refined the structure for both space groups using JANA software. The refined structures at 5~GPa and 12~GPa at 10~K, in the superconducting phase, are reported in SI \cite{SI}. The refinements yielded similar results for both space groups, with agreement factor values ($R_{obs}$) of 6.81 ($P2_1/m$) and 6.33 ($P2_1$) at 5~GPa, and 3.77 ($P2_1/m$)  and 3.76 ($P2_1$) at 12~GPa (see SI for details). Although $R_{obs}$ is marginally better for the $P2_1$ space group, the difference is within the margin of uncertainty, making it unreasonable to definitively assert this space group as the actual space group based on these grounds alone. To resolve this ambiguity, we performed geometry optimisation using DFT. However, similarly to XRD, it was impossible to conclude as no significant energy lowering was observed in the $P2_1$ group. We then examined the lattice dynamics using infrared and Raman spectroscopy, comparing the results with {\it ab-initio} calculations.

\paragraph{Lattice dynamics.}
\begin{figure}[h!]
\includegraphics[width=0.99\linewidth, angle=0]{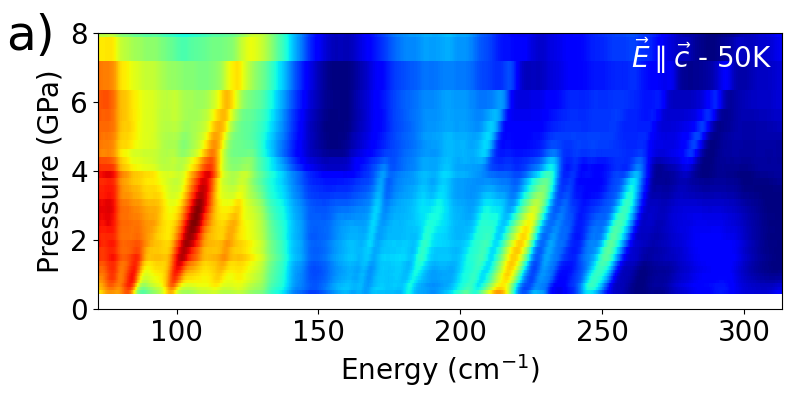}
\includegraphics[width=0.99\linewidth, angle=0]{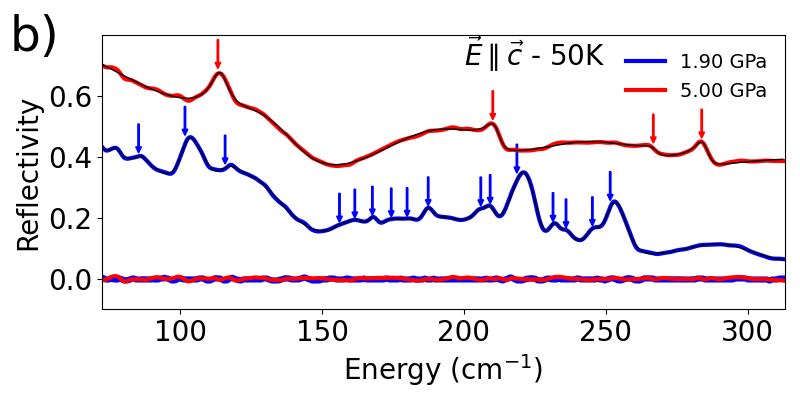}
\includegraphics[width=0.99\linewidth, angle=0]{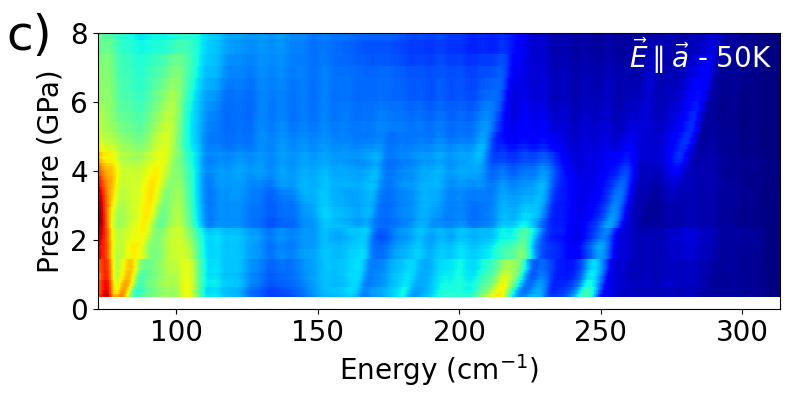}
\includegraphics[width=0.99\linewidth, angle=0]{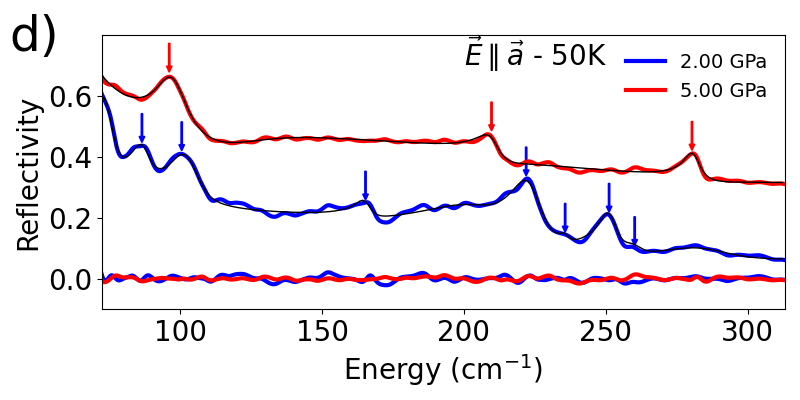}
\caption{(Color online) a and c : Reflectivity of \bfse\ as a function of energy and pressure for $\mathbf{E} \parallel \mathbf{c}$ ($B_{1u}$) and $\mathbf{E} \parallel \mathbf{a}$ ($B_{3u}$)  in $Cmcm$ setting, respectively. The value of the reflectivity at 310~$\rm cm^{-1}$ has been subtracted to remove the baseline shift due to progressive metallization. Panels b and d show selected reflectivity at 2 GPa (blue line) and 5 GPa (red line) for both polarizations. The fit of the reflectivity is represented by the black line, and the difference between the data and the fit is shown in the corresponding color. Arrows indicate the positions of the fitted phonons.}
\label{IR}
\end{figure}

\begin{figure}[h!]
\includegraphics[width=0.999\linewidth, angle=0]{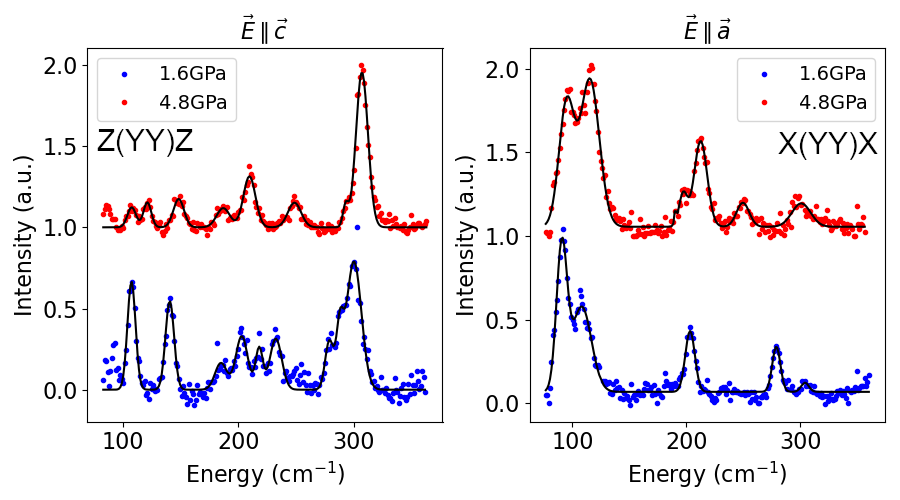}
\caption{(Color online) Raman spectra of \bfse\ at 1.6 (blue line) and 4.8 GPa (red line) at 300K for both polarizations of the incident light (left : $\mathbf{E} \parallel \mathbf{c}$ corresponding to Z(YY)Z; right $\mathbf{E} \parallel \mathbf{a}$ corresponding to X(YY)X with X,Y,Z along a,b,c in $Cmcm$ setting). They belong to $A_{g}$ irreducible representation in this configuration. The black line correspond to the fit with Gaussian functions.}
\label{Raman}
\end{figure}

We measured the pressure evolution of the IR reflectivity at 50\,K for two configurations of the incident polarization: with the electric field aligned along the $\mathbf a$ or $\mathbf c$ direction (in $Cmcm$ notation). 
In the centrosymmetric $Cmcm$ phase, infrared-active phonons transform according to the $B_{1u}$, $B_{2u}$, and $B_{3u}$ irreducible representations. Within the conventional $Cmcm$ setting, these modes are associated with electric fields polarized along $\mathbf c$ ($B_{1u}$), $\mathbf b$ ($B_{2u}$), and $\mathbf a$ ($B_{3u}$), respectively. As a consequence, our infrared measurements probe only the $B_{1u}$ modes for $\mathbf{E}\parallel\mathbf c$ and the $B_{3u}$ modes for $\mathbf{E}\parallel\mathbf a$, while the $B_{2u}$ modes polarized along $\mathbf b$ are not accessible in our experimental geometry.

At low pressure, we resolved the phonons previously published~\cite{Weseloh2022} and observed the expected hardening under pressure up to 4~GPa. It was associated with a decrease in amplitude likely due to the progressive metallization already reported~\cite{Ying2017}.

Above the 4~GPa transition, there is a substantial increase in the optical conductivity, resulting in an elevated baseline for the reflectivity, making it a challenge to discern the phonons. Consequently, only the most intense phonons remain visible and can be fitted. In Fig.~\ref{IR}a and c, we present a map of the reflectivity evolution as a function of pressure, to facilitate the visualization of the phonons and their pressure-induced changes. This approach helps distinguishing phonons from baseline oscillations. The phonons shift in energy with pressure, while the experimental artifacts, such as oscillations, remain at the same energy.

In Fig.~\ref{IR}b and d, we depict the reflectivity at 2 and 5~GPa for both polarizations, fitted with the standard Drude--Lorentz model, where the dielectric function is expressed as the sum of harmonic oscillators~\cite{Weseloh2022}. The extracted phonon frequencies are reported in Table~\ref{tab:phonon} for the 2~GPa and 5~GPa measurements.

At low temperature and high pressure, where the monoclinic symmetry is stabilized, the possible centrosymmetric space group is $P2_1/m$. In this case, infrared-active modes transform as $A_u$ and $B_u$. For an electric field polarized along $\mathbf c$ (defined with respect to the parent $Cmcm$ axes), the infrared-active modes belong to the $A_u$ representation, while for $\mathbf{E}\parallel\mathbf a$ they belong to the $B_u$ representation. Raman-active modes in the $P2_1/m$ group transform as $A_g$ and $B_g$. However, since our Raman measurements are performed in configurations where the incident and scattered light polarizations are parallel, only $A_g$ modes are expected to be observed.

We also performed Raman spectroscopy at ambient temperature. As noted in previous Raman scattering studies on these samples, the signal-to-noise ratio is particularly degraded for this family of systems~\cite{Popovic2015}, and it worsens at high pressure due to the increased metallicity. After several attempts, we successfully measured the Raman spectra for two configurations of polarization with both the incoming and outgoing electric fields aligned along the $\mathbf a$ and $\mathbf c$ directions, respectively, corresponding to the rungs and the legs of the ladders in $Cmcm$ notation.

In the centrosymmetric $Cmcm$ phase relevant at room temperature above 4~GPa, Raman-active phonons belong exclusively to the $A_g$ irreducible representation in our configuration. The mutual exclusion between Raman and infrared activities is lifted in the non-centrosymmetric $Pm$ phase, where phonons can be simultaneously Raman and infrared active. In this latter case, modes polarized along $\mathbf c$ transform as $A$, while modes polarized along $\mathbf a$ transform as $B$.

The Raman spectra at 1.6 and 4.8~GPa are presented in Fig.~\ref{Raman}, together with fits using Gaussian functions for each mode. The fitted frequencies, together with their pressure evolution, are reported in the Supplementary Information.

\paragraph{{\it Ab-initio} calculations.}
Phonons calculations were performed for both the $P2_1/m$ and $P2_1$ space groups. The structure was first relaxed under an applied pressure of 6 GPa, starting from the X-ray diffraction structural model (indeed, the 5 GPa calculations did not converge, most probably because the phase transition is close to this pressure in the DFT calculations).  The frequencies of the computed lattice dynamics suggest that both structures are dynamically stable. Table~\ref{tab:phonon} displays the DFT and experimental results in the irreducible representations (irreps) in which we have IR low temperature measurements for comparison. The $\mathbf{E} \parallel \mathbf{c}$ IR measurements correspond to the $A_u$ irrep in the $P2_1/m$ group and $A$ in the $P2_1$ group, while the $\mathbf{E} \parallel \mathbf{a}$ IR measurements correspond to the $B_u$ irrep in the $P2_1/m$ group and $B$ in the $P2_1$ group. 

\begin{table}[h!]
\centering
\caption{Frequencies (cm$^{-1}$) of optical phonons in the $P2_1/m$ and $P2_1$ space groups, obtained from DFT calculations and IR measurements at 50~K. Only the DFT modes with frequencies above 80~cm$^{-1}$ are listed. See SI for a complete description. The IR mode typed in bold red characters cannot be assigned to any computed one. In the $Cmcm$ setting, $c$ and $a$ are along the leg and the rung of the ladders. The star indicates a phonon visible on the IR raw data (see Fig. \ref{IR}) but that could not be properly fitted. This value is an estimation leading to larger uncertainty.}
\label{tab:phonon}
\begin{tabular}{ccc@{\;\;}cc} \\[-01ex]
  \multicolumn{2}{c}{$P2_1/m$} &  & \multicolumn{2}{c}{$P2_1$} \\[-1ex]
    \begin{tabular}[t]{cc}
        \hline \hline \rule{0pt}{2.5ex}{\hfill}
         DFT  & IR  \\ 
        A$_u$    & $\mathbf{E} \parallel \mathbf{c}$ \\ 
        \hline  
        81  &       \\
        90  &       \\
        104 &       \\
        122 & 113   \\
        157 &       \\
        181 &       \\
        209 & 210   \\
        215 &       \\
        226 &    \\
        233 &    \\
        252 & 267\\
            & \textcolor{red}{\bf 284}   \\
        \hline \hline
    \end{tabular}
 & 
    \begin{tabular}[t]{cr}
        \hline \hline  \rule{0pt}{2.5ex}{\hfill}
         DFT  & IR  \\ 
        B$_u$    & $\mathbf{E} \parallel \mathbf{a}$ \\ 
        \hline
        87  &  \\
        93  &  \\
        94  &  96 \\
        106 &  \\
        132 &  \\
        139 &  \\
        151 &  \\
        184 &  \\
        201	& 210 \\
        233	&  \\        
        251	&  \\
        269 &*266 \\
        280 & 280 \\
        298 &  \\
        320 &  \\
        \hline \hline
    \end{tabular}
 & & %
    \begin{tabular}[t]{cc}
        \hline \hline \rule{0pt}{2.5ex}{\hfill}
      DFT  & IR  \\ 
        A     & $\mathbf{E} \parallel \mathbf{c}$ \\ 
        \hline
           80  &      \\
           85  &      \\
           90  &      \\
           94  &      \\
           101 &      \\
           104 &      \\
           109 & 113  \\
           122 &      \\
           124 &      \\
           132 &      \\
           142 &      \\
           156 &      \\
           179 &      \\
           187 &      \\
           204 &      \\
           208 & 210  \\
           216 &      \\
           226 &      \\
           232 &      \\
           233 &      \\
           252 &      \\
           255 &      \\
           273 & 267  \\
           275 &      \\
           292 & 284  \\
           315 &      \\
           \hline  \hline
    \end{tabular}
& 
    \begin{tabular}[t]{cr}
        \hline \hline \rule{0pt}{2.5ex}{\hfill}
         DFT  & IR  \\ 
        B    & $\mathbf{E} \parallel \mathbf{a}$  \\ 
        \hline
           87  &    \\
           89  &    \\
           92  &    \\
           95  & 96 \\
           99  &    \\
           106 &    \\
           122 &    \\
           130 &    \\
           140 &    \\
           151 &    \\
           157 &    \\
           181 &    \\
           184 &    \\
           201 &    \\
           210 & 210\\
           217 &    \\
           227 &    \\
           233 &    \\
           251 &    \\
           268 &*266\\
           279 & 280\\
           299 &    \\
           320 &    \\
           \hline  \hline
    \end{tabular} 
\end{tabular}
\end{table}

As we can see immediately, in the $P2_1/m$ group the phonon mode at $284~\rm cm^{-1}$ for $\mathbf{E} \parallel \mathbf{c}$ cannot be assigned to any computed mode, while all modes can easily be assigned in the $P2_1$ group. In fact this mode is close in energy to two $A_g$ mode of the $P2_1/m$ group (see SI), in agreement with the loss of the inversion center. Moreover, although it is less critical, the mode at $267~\rm cm^{-1}$ for $\mathbf{E} \parallel \mathbf{c}$ deviates,  from the last calculated phonon at $252~\rm cm^{-1}$, further than the typical uncertainty, which is on the order of $8~\rm cm^{-1}$. Finally, despite the fact that the Raman spectra have been measured at 300~K and thus should somewhat deviate from our 0~K calculations, one can see that the mode at 197~cm$^{-1}$ for $\mathbf{E} \parallel \mathbf{a}$  cannot be associated to any computed $A_g$ modes (see Fig. \ref{Raman} and SI).

From all these arguments we can reasonably conclude that the high-pressure low-temperature phase crystallizes in the $P2_1$ group.  

\begin{figure}[h!]
  \centering
  \resizebox{\linewidth}{!}{\includegraphics{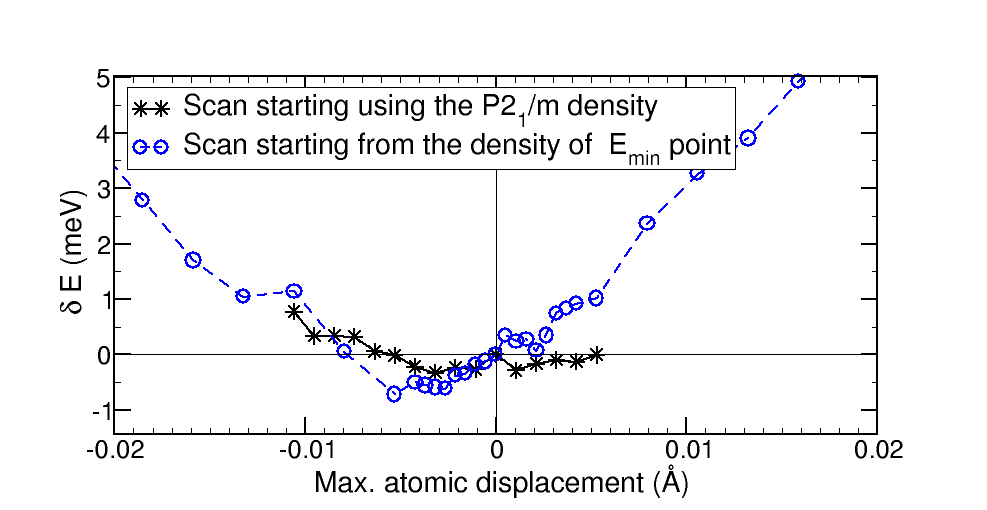}}
  \caption{Energy difference along the soft $P2_1/m$ to $P2_1$ phonons
    mode. The black curve uses the $P2_1/m$ density as starting point for the
    self-consistent process, while the blue one uses the density of the
    minimum energy point of the black curve.}
  \label{fig:deltaE}
\end{figure}

Since the high-pressure space group, $P2_1$, is not a subgroup of the ambient-pressure one, $Pm$, the transition is expected to be first-order. 
Of course, we cannot exclude the theoretical possibility of an intermediate phase with $P2_1/m$ symmetry in a narrow pressure range. In such a scenario, the structural evolution would proceed via two successive second-order transitions. 
The first order transition senario is supported by the observation of a slight coexistence of phonon modes around 4\;GPa in the $\mathbf{E} \parallel \mathbf{a}$ polarization configuration. However, this observation could also be attributed to a pressure gradient within the sample, leading to spatial phase separation and the coexistence of domains corresponding to different structures.
The two transitions senario is supported by the flatness of the energy surface between the $P2_1/m$ and $P2_1$ structures (see Fig.~\ref{fig:deltaE}). Although the distorted $P2_1$ structure clearly corresponds to an energy minimum, the distortion itself remains weak, with atomic displacements on the order of $\sim 0.005$~\AA, comparable to those found in spin-driven multiferroics. Moreover, even for displacements as large as 0.03~\AA, the energy difference remains within a few meV. This suggests that a transition pathway involving two successive second-order transitions is, in principle, possible. 
Nevertheless, we consider this explanation unlikely, as the coexistence occurs over a pressure range of approximately 0.5~GPa, which exceeds the expected pressure gradient in the cell at such low pressure. In conclusion, although a multi-step transition pathway cannot be ruled out, the most likely scenario remains that the structural transition is first order.

\paragraph{Conclusion.}
Through a combined experimental and theoretical approach, including X-ray diffraction, infrared and Raman spectroscopies, and {\it ab-initio} calculations, we have determined the atomic structure of the high-pressure, low-temperature phase of \bfse\ in which superconductivity emerges. Our results demonstrate that this phase crystallizes in the polar $P2_1$ space group, establishing \bfse\ as a rare example of a quasi-one-dimensional non-centrosymmetric superconductor\cite{Bao2015, Gosar2023}.
The absence of inversion symmetry allows, in the presence of antisymmetric spin--orbit coupling, for the mixing of spin-singlet and spin-triplet pairing channels, among other exotic phenomena~\cite{Yip2014, GorkovRashba2001,Samokhin2009,Smidman2017,Kneidinger2015}. In Fe-based systems, where the intrinsic spin--orbit coupling is moderate, such mixing is expected to be limited; nevertheless, broken inversion symmetry qualitatively alters the allowed superconducting pairing states.
These findings provide a well-defined structural foundation for future investigations of superconductivity in reduced dimensionality and highlight \bfse\ as a particularly promising platform for exploring the interplay between lattice symmetry, magnetism, and superconductivity in iron-based materials.

\paragraph{Aknowledgment.}
We thank SOLEIL for synchrotron beam time (Proposal 20210631 and 20220451). V.B. acknowledges the MORPHEUS platform at the Laboratoire de Physique des Solides. This work was financially supported by the ANR COCOM 20-CE30-0029, the France 2030 programme "ANR-11-IDEX-0003" via Integrative Institute of Materials from Paris-Saclay University - 2IM@UPSaclay and by the CSC scholarship (No. 201806830111). M.-B.L., S.D. and G.G. aknowledge the IDRIS high-performance computer center for the calculation time under the GENCI project n$^\circ$0801842.

\bibliography{BFS.bib}
\end{document}



\title{\bfse : a quasi-unidimensional non-centrosymmetric superconductor \\ Supplemental Information}

\author{S. Deng$^{*}$}
\affiliation{Institut Laue Langevin, 38000 Grenoble, France}

\author{A. Roll$^{*}$}
\affiliation{Universit\'e Paris-Saclay, CNRS, Laboratoire de Physique des Solides, 91405, Orsay, France.}
\affiliation{Universit\'e Paris-Saclay, CNRS-CEA, Laboratoire L\'eon Brillouin, 91191, Gif sur Yvette, France}

\author{W. G. Zheng$^{*}$}
\affiliation{Universit\'e Paris-Saclay, CNRS, Laboratoire de Physique des Solides, 91405, Orsay, France.}

\author{G. Giri}
\affiliation{Institut Laue Langevin, 38000 Grenoble, France}
\affiliation{Current address: Physics Dept., VIT Bhopal University, Bhopal, India}

\author{T. Vasina}
\affiliation{Institut Laue Langevin, 38000 Grenoble, France}

\author{D. Bounoua}
\affiliation{Universit\'e Paris-Saclay, CNRS-CEA, Laboratoire L\'eon Brillouin, 91191, Gif sur Yvette, France}

\author{P. Fertey}
\affiliation{Synchrotron SOLEIL, L\'\ Orme des Merisiers, Saint Aubin BP 48, 91192, Gif-sur-Yvette, France}

\author {M. Verseils}
\affiliation{Synchrotron SOLEIL, L\'\ Orme des Merisiers, Saint Aubin BP 48, 91192, Gif-sur-Yvette, France}

\author {C. Bellin}
\affiliation{Universit\'e Pierre et Marie Curie, IMPMC, CNRS UMR7590, 4 Place Jussieu, 75005 Paris, France}

\author{A. Forget}
\affiliation{Universit\'e Paris-Saclay, CEA, CNRS, SPEC, 91191, Gif-sur-Yvette, France.}

\author{D. Colson}
\affiliation{Universit\'e Paris-Saclay, CEA, CNRS, SPEC, 91191, Gif-sur-Yvette, France.}

\author{P. Foury-Leylekian}
\affiliation{Universit\'e Paris-Saclay, CNRS, Laboratoire de Physique des Solides, 91405, Orsay, France.}

\author {M.B. Lepetit}
\affiliation{Institut N\'eel, CNRS, 38042 Grenoble, France}
\affiliation{Institut Laue Langevin, 38000 Grenoble, France}

\author{V. Bal\'edent}
\affiliation{Universit\'e Paris-Saclay, CNRS, Laboratoire de Physique des Solides, 91405, Orsay, France.}
\affiliation{Institut universitaire de France (IUF)}
\email [Corresponding author: ] {victor.baledent@universite-paris-saclay.fr}

\affiliation{$^{*}$These authors contributed equally to this work.}

\date{\today}

\maketitle

\section{Symmetry analysis}
The general reflection conditions in the  $Cmcm$ space group are provided in Table~\ref{tab:RC_Cmcm}
\begin{table}[h!]
\centering
\caption{Reflection conditions in the $Cmcm$ space group.}
    \begin{tabular}{l@{ : }l@{\qquad}l@{ : }l@{\qquad}l@{ : }l}
    \hline \hline 
    hkl& h+k=2n \\
    0kl& k=2n & h0l& h,l=2n & hk0& h+k=2n \\
    h00& h=2n & 0k0& k=2n & 00l& l=2n \\
    \hline \hline
    \end{tabular} \hfill \\
\label{tab:RC_Cmcm}
\end{table}

As the $Cmcm$ space group is $C$-centered, its unit cell therefore contains two primitive unit cells. The $Cmcm$ conventional unit cell and the primitive one are related as follow \\

\begin{minipage}{0.3\linewidth}
\begin{eqnarray}
    {\bf a_P} &=& \frac{1}{2}({\bf a_C} + {\bf b_C}) \\
    {\bf b_P} &=& \frac{1}{2}(-{\bf a_C} + {\bf b_C}) \\
    {\bf c_P} &=& {\bf c_C}
\end{eqnarray}
\end{minipage}
\begin{minipage}{0.3\linewidth}
\begin{eqnarray}
    {\bf a^*_P} &=& {\bf a^*_C} + {\bf b^*_C} \\
    {\bf b^*_P} &=& -{\bf a^*_C} + {\bf b^*_C} \\
    {\bf c^*_P} &=& {\bf c^*_C}
\end{eqnarray}
\end{minipage}
\begin{minipage}{0.3\linewidth}
\begin{eqnarray}
    { h_C} &=& h_P-k_P \\
    { k_C} &=& h_P+k_P \\
    { l_C} &=& l_P
\end{eqnarray}
\end{minipage} \hfill \\

\noindent 
where the subscripts $P$ refers to the primitive unit cell and $C$ to the conventional. \\

Let us analyze what the diffraction pattern of \bfse\ at 10\;K and 5 or 10\;GPa tells us (see Fig.1 of main paper). 
\begin{itemize}
    \item We observe intensity on the $h_C, k_C, l_C: \; h_C+k_C = 2n+1$  peaks, including on the $l_C=0$ ones. \\
    As the $C$-centering implies $ h_C+k_C = 2h_P, \quad h_P \in \mathbb{Z}$, the partial translation associated to the $C$-centering is lost. The primitive unit cell of the true space group is therefore the conventional $Cmcm$ one, which is twice the primitive one, and the true group should be a  $Cmcm$ subgroup of $k$-index=2.

    \item  We observe intensity on the $h_C, 0, l_C: \; l_C = 2n+1$  peaks.  \\
    As the $c$ symmetry operation $\{ m_{010} | 0 0 \frac{1}{2} \}$ implies   $h_C, 0, l_C: \; l_C = 2n$, this symmetry operation does not belong to the true space group. 

    \item When it comes to the  $(0,0,l_C)$ reflexions, there is no sign of intensity on any $l_C=2n+1$ peak. Even though not seen any intensity is not a strong proof, one can safely suppose that the $2_1$ $\{ 2_{001} | 0 0 \frac{1}{2} \}$ symmetry operation is present in the true space group.  

    \item The diffraction pattern pictured in Fig.1 of the main paper clearly shows that the true structure should be monoclinic and not orthorhombic, with a unique axis along the ${\bf c}$ $Cmcm$ direction,. 
\end{itemize}

Taking off the $C$-centering and the $c: \{ m_{010} | 0 0 \frac{1}{2} \}$ mirror from the generators of $Cmcm$,  only five subgroups keep the $2_1: \, \{ 2_{001} | 0 0 \nicefrac{1}{2} \}$ screw axis. Out of them  
only two are monoclinic, namely 
\begin{itemize}
    \item $P2_1/m$ with symmetry operations $\{ 1 | 0 \}, \{ 2_{001} | 0 0 \frac{1}{2} \}, \{ -1 | 0 \}, \{ m_{001} | 0 0 \frac{1}{2} \}$~;
    \item   and $P2_1/c$  with symmetry operations $\{ 1 | 0 \}, \{ 2_{001} | 0 0 \frac{1}{2} \}, \{ -1 | \frac{1}{2}\frac{1}{2} 0 \}, \{ m_{001} | \frac{1}{2}\frac{1}{2}\frac{1}{2}\}$.
\end{itemize}
The $c: \{m_{001} | \frac{1}{2}\frac{1}{2}\frac{1}{2}\}$ mirror implies the following reflection conditions $h_C,k_C,0:\; k_C+k_C=2n$. As we observe intensity on the $h_C,k_C,0:\; k_C+k_C+2n+1$ peaks, the only possible space groups for \bfse\ uneder pressure and at low temperature are $P2_1/m$ and its subgroups.


\section{X-ray Diffraction}
For x-ray diffraction, a diamond anvil cell (DAC) with 600\;$\mu$m culets was used, with a sample chamber of 300\;$\mu$m of diameter and 50\;$\mu$m thick. Several small crystals (about 40\;$\mu$m of size) were loaded together with a small amount of gold powder and a ruby chip to measure the pressure. The presence of several crystals increases the chances of having a good quality crystal that resists pressure. Helium was used as pressure transmitting medium. Low temperature X-ray diffraction measurements were performed using a newly developed helium circulation cryostat specially designed for diamond anvil cells, which allows for cell rotation at low temperatures, while maintaining the ability to adjust pressure \textit{in situ}. This rotation axis, in conjunction with the cryostat's rotation around a vertical axis, facilitates the collection of a large number of Bragg peaks, thereby improving the completeness of the data sets and the precision of structural refinements. This specialized cryostat is jointly operated by the Laboratoire de Physique des Solides (Orsay, France) and the CRISTAL beamline at the SOLEIL synchrotron (Gif-sur-Yvette, France).  The pressure inside the diamond anvil cell was measured using the ruby fluorescence technique, yielding a typical 5\% relative accuracy~\cite{Shen2020}. The pressure measurement was confirmed by the pressure deduced from the gold equation of state and the cell parameter determination of the gold powder sample, also loaded in the DAC chamber. Data collections were conducted at a wavelength of $\lambda = 0.4175$\;\AA, at the CRISTAL beamline using a 2D detector and processed with the CrysAlis Pro~\cite{crysalis}.

Table~\ref{tab:Robs} provides the agreement factor values of the refinements ($R(obs)$), as well as the number of peaks used at each measured point of the pressure-temperature phase diagram. 
\begin{table}[h!]
\caption{Value of the agreement factors (R(obs)) in \% for each refinement for selected temperature and pressure. The number in parenthesis corresponds to the number of independent peaks taken into account, with an intensity over noise ratio greater than 3.}      \label{tab:Robs}    
\begin{tabular}{c|c|cccccc}
\hline \hline \rule{0pt}{2.5ex}{\hfill}
Temperature & Pressure & $Pnma$ & $Pmn2_1$ & $Pm$ & $Cmcm$ & $P2_1/m$ & $P2_1$ \\
\hline \rule{0pt}{2.5ex}{\hfill}
\multirow{2}{*}{ 300 K } & 0 GPa & 5.46 (411) & 2.18 (619) & 1.86 (647) & & & \\
 & 12 GPa & 5.47 (625) & 4.54 (761) & 4.63 (1069) & 4.38 (516) & 17.52 (1145)  & 15.87 (1807) \\
\hline \rule{0pt}{2.5ex}{\hfill}
\multirow{3}{*}{ 10 K } & 0 GPa & 7.27 (672) & 7.08 (936) & 7.8 (1419) & & & \\
 & 5 GPa & & & & 31.14 (531) & 6.81 (1401) & 6.33 (1859) \\ 
 & 12 GPa & & & & & 3.77 (403) & 3.76 (690) \\ \hline \hline
\end{tabular}

\end{table}

\begin{figure}[h!]
\caption{Calculated intensity versus measured intensity at 10\;K and at 5\;GPa using $P2_1/m$ (a), $P2_1$ (b) and $Cmcm$ (c) space groups, and at 12 GPa using $P2_1/m$ (d) and $P2_1$ (e) space groups.}
\includegraphics[width=0.9\linewidth, angle=0]{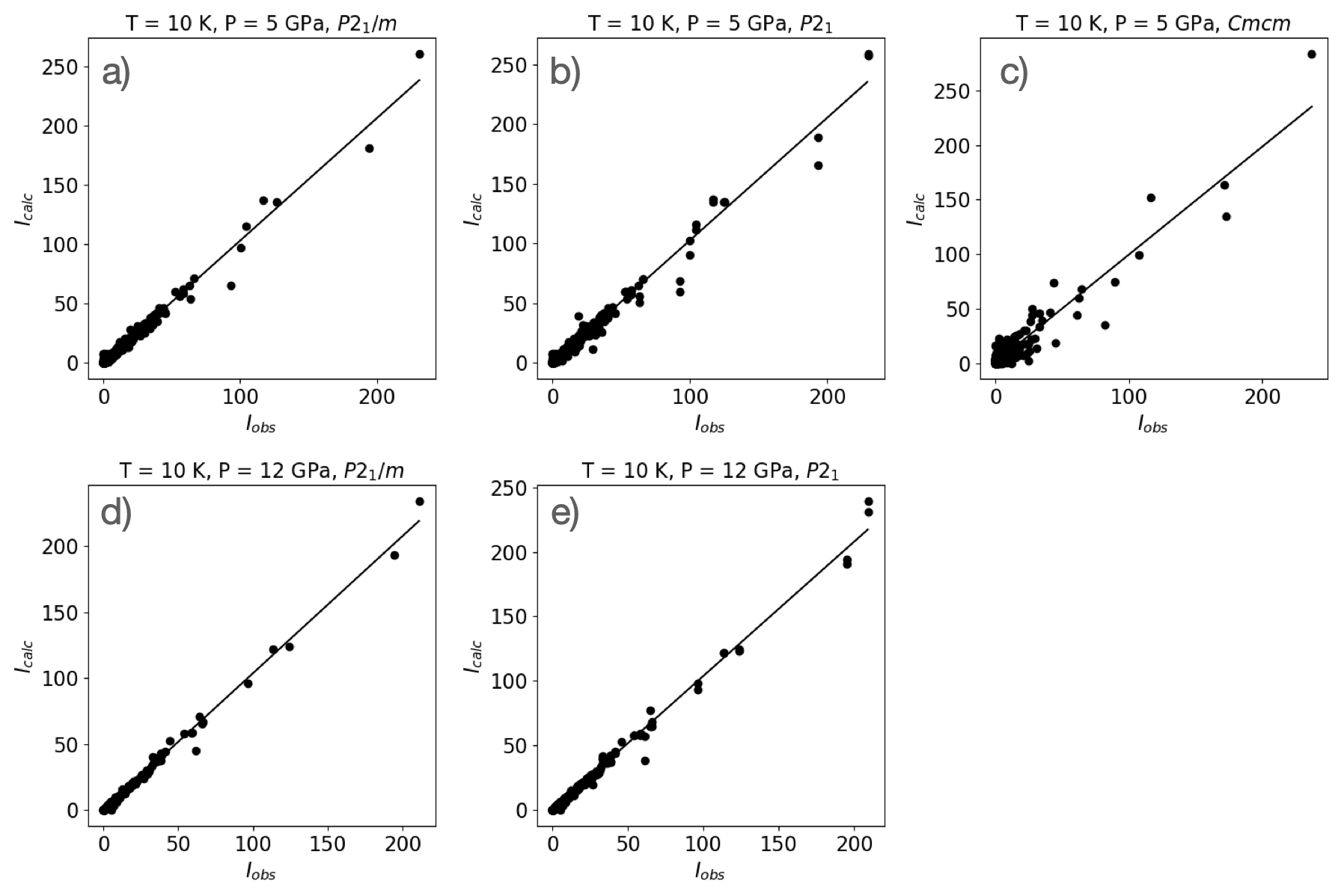}
\label{Fig_SI_XRD}
\end{figure}

For the high-pressure, low-temperature data points (10\;K), Fig.~\ref{Fig_SI_XRD} compares the calculated versus measured intensities, for the two monoclinic space groups $P2_1/m$ and $P2_1$, and the 5 and 12\;GPa pressure values, for which similar refinement quality was obtained. For comparison the same plot is made for $Cmcm$ space group at 5\;GPa and 10\;K.

\begin{table}[h!]
{\bf $P2_1$ at 10\;K and 5\;GPa }
\caption{Atomic positions of \bfse\ at 10\;K and 5\;GPa in the $P2_1$ space group (R=6.33\%). The lattice parameters are a=7.211(1)\;\AA, b=5.3402(1)\;\AA, c=14.330(1)\;\AA, and $\beta$=102.7(1)\;\degree.}
     {\small
       \begin{tabular}{ldddd}
       \hline \hline
       Element & $x$ & $y$ & $z$ & $U$ \\ \hline 
       Ba1 & 0.7539(3) & 0.2501(6) & 0.07555(11) & 0.0210(10) \\
       Ba2 & 1.3101(3) & 0.2493(5) & 0.59099(12) & 0.0222(11) \\
       Se1 & 0.2120(5) & 0.2508(10) & 0.04479(18) & 0.0198(16) \\
       Se2 & 0.4335(5) & 0.2510(9) & 0.83419(20) & 0.0197(17) \\
       Se3 & 0.9592(5) & 0.2487(9) & 0.81804(18) & 0.0159(16) \\
       Se4 & 1.1579(5) & 0.7534(10) & 0.69454(19) & 0.0190(16) \\
       Se5 & 0.8226(5) & 0.2470(11) & 0.54096(19) & 0.0235(17) \\
       Se6 & 0.5874(5) & 0.7504(9) & 0.6593(2) & 0.0237(18) \\
       Fe1.1 & 0.8839(16) & 0.5060(10) & 0.6792(7) & 0.002(5) \\
       Fe1.2 & 0.125(2) & 0.4934(10) & 0.3227(9) & 0.045(7) \\
       Fe2.1 & 0.6866(19) & 0.5112(9) & 0.8114(7) & 0.020(5) \\
       Fe2.2 & 0.315(2) & 0.4899(10) & 0.1871(8) & 0.033(7) \\
       \hline \hline
       \end{tabular}
       }
 \end{table}

 \bigskip

\begin{table}[h!]
{\bf $P2_1$ at 10\;K and 12\;GPa}
\caption{Atomic positions of \bfse\ at 10\;K and 12\;GPa, i.e. in the superconducting phase, in the $P2_1$ space group (R=3.76\%). The lattice parameters are a=6.752(1)\;\AA, b=5.227(1)\;\AA, c=13.866(1)\;\AA, and $\beta$=103.4(1)\;\degree.}
\begin{tabular}{cdddd}
\hline \hline
Element & $x$ & $y$ & $z$ & $U$ \\ \hline 
Ba & 0.7443(5) & 0.2506(8) & 0.07123(17) & 0.0183(19) \\
Ba & 1.3254(5) & 0.2500(7) & 0.58658(18) & 0.0174(18) \\
Fe & 0.894(3) & 0.5002(16) & 0.6758(14) & 0.004(10) \\
Fe & 0.104(4) & 0.501(2) & 0.3226(15) & 0.032(12) \\
Fe & 0.677(4) & 0.5058(19) & 0.8072(14) & 0.012(12) \\
Fe & 0.303(3) & 0.4939(16) & 0.1865(15) & 0.043(11) \\
Se & 0.2147(8) & 0.2535(14) & 0.0461(3) & 0.015(3) \\
Se & 0.4279(8) & 0.2502(13) & 0.8345(3) & 0.017(3) \\
Se & 0.9702(8) & 0.2501(15) & 0.8191(3) & 0.015(3) \\
Se & 1.1778(8) & 0.7474(15) & 0.6964(3) & 0.013(3) \\
Se & 0.8208(8) & 0.2464(13) & 0.5353(3) & 0.009(3) \\
Se & 0.6089(8) & 0.7501(13) & 0.6611(3) & 0.017(3) \\
\hline \hline
\end{tabular}
\end{table}

\begin{table}[h!]
\hfill \\[2ex]
{\bf $Cmcm$ at 300\;K and 12\;GPa} 
\caption{Atomic positions of \bfse\ at 300\;K and 12\;GPa in the $Cmcm$ space group (R=4.38\%). The lattice parameters are a=8.704(1)\;\AA, b=10.453(1)\;\AA, and c=5.260(1)\;\AA.}
\begin{tabular}{cdddd}
\hline \hline
Element & $x$ & $y$ & $z$ & $U$ \\ \hline 
Ba & 0.5        & 0.17152(13) & 0.25 & 0.0173(5) \\
Se & 0.5        & 0.6324(2)   & 0.25 & 0.0141(7) \\
Se & 0.20054(8) & 0.37143(17) & 0.25 & 0.0189(6) \\
Fe & 0.34977(9) & 0.5         & 0    & 0.0153(8) \\
\hline \hline
\end{tabular}
\end{table}

\begin{figure}[h]
\caption{(Color online)(a) Schematic temperature-pressure phase diagram for the structure of \bfse. (b-d) Structures of \bfse\ in the different space groups obtained in the phase diagramm and their respective unit cell for comparison : $Pm$ (a), $Cmcm$ (b) and $P2_1$ (c).}
\includegraphics[width=0.6\linewidth, angle=0]{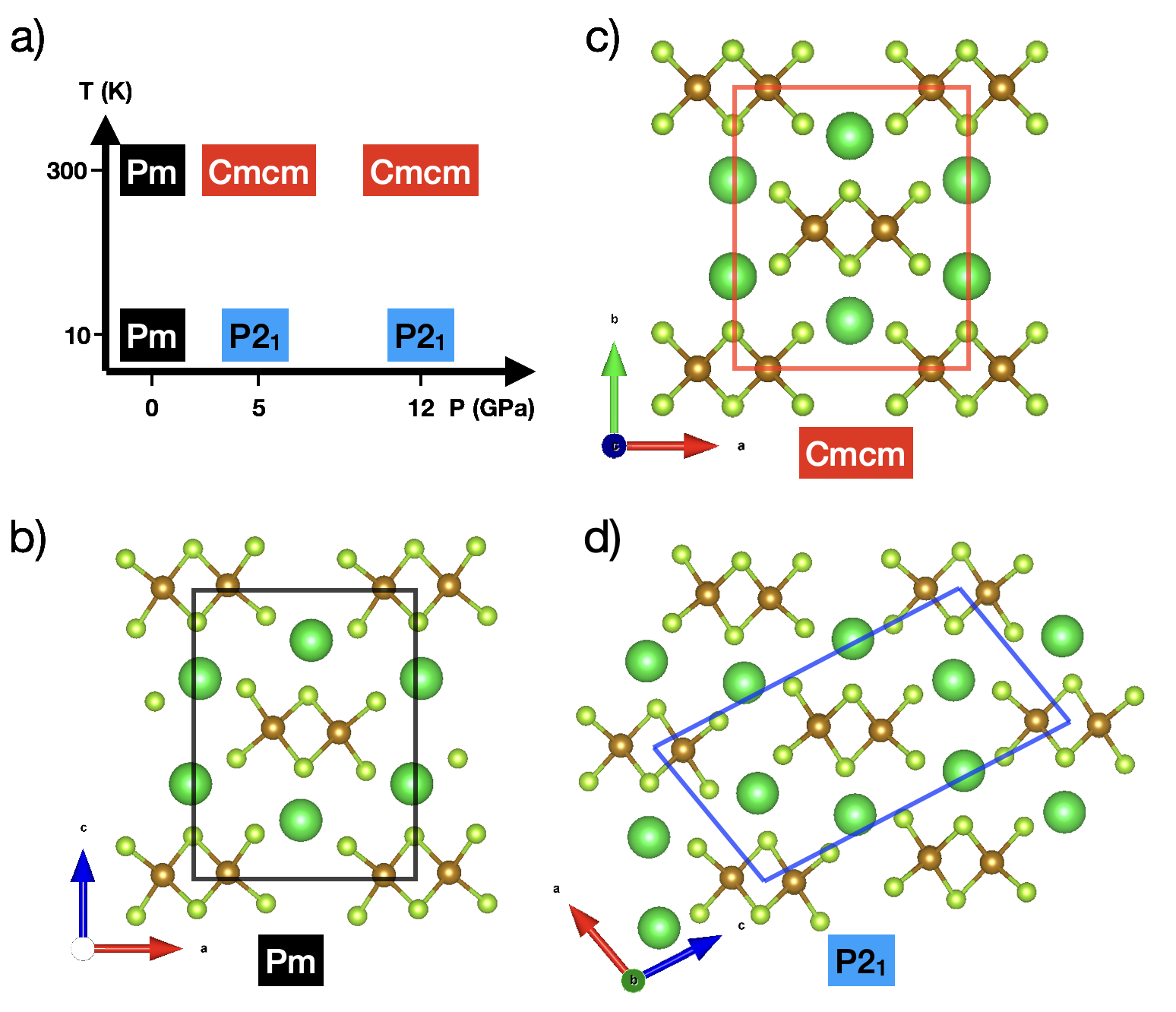}
\label{Fig_SI_Raman}
\end{figure}

\newpage 
\clearpage

\section{Infrared and Raman Spectroscopy}
Raman scattering experiments were performed at Institut de minéralogie, de physique des matériaux et de cosmochimie (IMPMC, Sorbonne University, France) with a 534\;nm laser source, with a fixed power of 15\;mW in a backscattering geometry. \bfse\ single crystal was loaded in a diamond anvil cell using diamonds of 1\;mm culet size and an inox gasket drilled with a 0.5\;mm hole. Si oil was used as a pressure transmitting medium as it is hydrostatic at room temperature up to 10\;GPa~\cite{Tateiwa2009}. The pressure inside the diamond anvil cell was measured using the ruby fluorescence technique, yielding a typical 5\% relative accuracy~\cite{Shen2020}. The orientation of the sample was the checked {\it in situ} at Laboratoire de Physique des Solides on the high pressure diffraction setup. A third-order polynomial background has been removed from the raw data, coming essentially from the tail of the elastic line. All data are presented in Fig.~\ref{Fig_SI_Raman} and fitted frequencies for 1.9\;GPa and 4.8\;GPa are reported in Table~\ref{tab:IR}.

Infrared spectroscopy measurements were carried out at the AILES beamline of the SOLEIL synchrotron, with a Bruker IFS125 Michelson interferometer~\cite{Roy2006} equipped with a circulating Helium cryostat specific for Diamond Anvil Cells, a 4.2\;K bolometer, and a 6\;$\mu$m beam splitter for a resolution of 2\;cm$^{-1}$. Two different cells were prepared with two different samples which orientation was checked {\it in situ} at Laboratoire de Physique des Solides on the high pressure diffraction setup. The pressure transmitting medium used was polyethylene, well suited for infrared spectroscopy~\cite{Celeste2019}. The pressure inside the diamond anvil cell was also measured using the ruby fluorescence technique, yielding a typical 5\% relative accuracy~\cite{Shen2020}. For each crystal, the reflectivity was recorded for several temperatures from 10 to 300 K. The absolute reflectivity of the sample was obtained by using the gasket reflectivity as reference. As a consequence, each reflectivity spectra is the ratio between the reflected intensities on the sample and on the reflectivity from the gasket.

\begin{figure}[h]
\caption{(Color online) Raman spectra of \bfse\ at four different pressure points for both polarizations of the incident and scattered light along $c$ and $a$ in $Cmcm$ setting: along the leg and the rung of the ladder respectively. The black line correspond to the fit with Gaussian functions.}
\includegraphics[width=0.7\linewidth, angle=0]{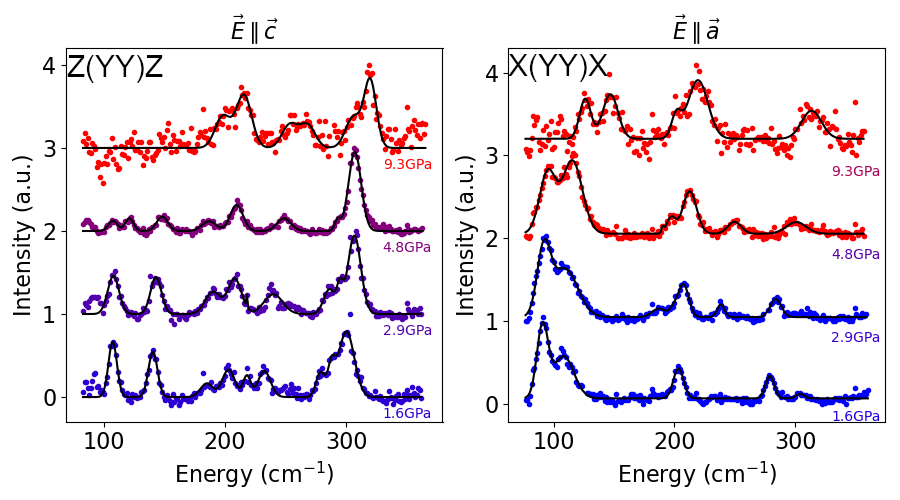}
\label{Fig_SI_Raman}
\end{figure}
  
\begin{table*}[h!]
\caption{Fitted phonons frequencies (in $cm^{-1}$) from infrared reflectivity Raman spectra for both polarization configuration $\mathbf{E} \parallel \mathbf{c}$ and $\mathbf{E} \parallel \mathbf{a}$ in the low pressure phase ( 1.6, 1.9 or 2\;GPa) and high pressure phase (4.8 or 5\;GPa) in the $Cmcm$ setting ($c$ and $a$ are along the leg and the rung of the ladder respectively). Error bars from the fit are given within parenthesis. The star indicates phonon visible on the IR raw data but could not be fitted. Its value is an estimation leading to larger uncertainty.}
\centering
\begin{tabular}{rr@{\quad}rr@{\qquad}rr@{\quad}rr}
  \hline\hline
  \multicolumn{4}{c}{IR (T = 50 K)} & \multicolumn{4}{c}{Raman (T = 300 K) }
                                      \rule{0pt}{3ex}{\hfill}
 \\[1ex]  \hline
  \multicolumn{2}{c}{$\mathbf{E} \parallel  \mathbf{c}$}
                                   & \multicolumn{2}{c}{$\mathbf{E} \parallel  \mathbf{a}$}
                                   & \multicolumn{2}{c}{Y(ZZ)Y ($\mathbf{E} \parallel  \mathbf{c}$)}
                                   & \multicolumn{2}{c}{X(ZZ)X ($\mathbf{E} \parallel  \mathbf{a}$)}\\
   \rule{0pt}{2ex}{\hfill}
  1.9~GPa & 5.0~GPa & 2.0~GPa & 5.0~GPa & 1.6GPa & 4.8~GPa & 1.6~GPa & 4.8~GPa \\
    \hline
86(3)  & 113(3) & 87(3)   & 96(3)   & 107(1) & 108(1) & 91(1) & 95(1) \\
102(3) & 210(3) & 101(8)  & 210(3)  & 140(1) & 121(1) & 108(1) & 116(1) \\
116(3) & 267(3) & 165(3)  & *266(8) & 185(1) & 148(1) & 203(1) & 190(1) \\
156(3) & 284(3) & *187(8) & 280(3)  & 203(1) & 187(1) & 279(1) & 197(1) \\
162(3) &        & 222(3)  &         & 218(1) & 210(1) & 304(1) & 213(1) \\
168(3) &        & 236(3)  &         & 232(1) & 249(1) &        & 250(1) \\
174(3) &        & 251(3)  &         & 279(1) & 293(1) &        & 301(1) \\
180(3) &        & 260(3)  &         & 288(1) & 307(1) &       & \\
187(3) &        &         &         & 300(1) &        &       & \\
206(3) &        &         &         &        &        &       & \\
209(3) &        &         &         &        &        &       & \\
219(3) &        &         &         &        &        &       & \\
231(3) &        &         &         &        &        &       & \\
236(3) &        &         &         &        &        &       & \\
245(3) &        &         &         &        &        &       & \\
252(3) &        &         &         &        &        &       & \\ \hline\hline
\end{tabular} 
\label{tab:IR}
\end{table*}

\newpage
\clearpage
\section{DFT calculations \vspace*{-2ex}} 
The DFT calculations were done using the CRYSTAL23 code~\cite{CRYSTAL23} that uses atomic gaussian basis set. The locality of the later allows the user of hybrid functionals at low additional computational cost ($\sim 20\%$).  The CRYSTAL code also allows a thorough symmetry control as all 230 space group are explicitly coded. Geometry optimizations are done by only using the symmetry allowed distortions of the space group specified in the inputs. Similarly, the phonons modes are computed in the irreps of the specified group.

In order to best account for the electronic correlation at the Fermi level, we used the hybrid B3LYP functional~\cite{B3LYP}.  An atomic gaussian basis of $3\zeta+P$ quality was used for the Fe and Se atoms~\cite{pob-TZVP-rev2}. The Ba atoms were represented using a relativistic core pseudo-potential of the Stuttgard group~\cite{PseudoR13} and the associated basis set where the diffuse functions exponents were taken as 1.2 as recommended by Scuseria {\it et al}~\cite{HSE}.  The shrinking factors were set to $8\times 12 \times 4$ resulting in a sampling net of 126 $\bf k$-points. The stripe spin ordering found in neutron scattering experiments of \bfse\ under pressure~\cite{Zheng2022} was used in all calculations.

\begin{table}[h!]
   \begin{minipage}[t]{0.45\linewidth}
     \caption{Transverse phonons modes computed using DFT (B3LYP functional) in the $P2_1$ group.\\[1ex]} 
     \label{Tab:DFT_P21}
     \begin{tabular}{r@{\quad}r}
      \hline\hline
      $A$  irrep & $B$  irrep \\
      \hline
      18      &         \\
              &  35     \\
      40      &  38     \\
      48      &         \\
      49      &         \\
      55      &         \\
      57      &  61     \\
      64      &  65     \\
      68      &  69     \\
              &  70     \\
              &  74     \\
              &  75     \\
      76      &  76     \\     
      80      &         \\
      85      & 87      \\
      90      & 89      \\
              & 92      \\
      94      & 95      \\
      101     & 99      \\
      104     & 106     \\
      109     &         \\
      122     & 122     \\
      124     &         \\
      132     & 130     \\
      142     & 140     \\
              & 151     \\
      156     & 157     \\
      179     & 181     \\
      182     & 184     \\
      204     & 201     \\
      208     & 210     \\
      216     & 217     \\
      226     & 227     \\
      232     &         \\
      233     & 233     \\
      252     & 251     \\
      255     & 252     \\
              & 268     \\
      275     & 279     \\
      292     &         \\
              & 299     \\
      315     &         \\
              & 320     \\
      \hline \hline
    \end{tabular}
   \end{minipage} \hfill
      \begin{minipage}[t]{0.45\linewidth}
     \caption{Transverse phonons modes computed using DFT (B3LYP functional) in the $P2_1/m$ group. \\[1ex]}
     \label{Tab:DFT_P21sm}
     \begin{tabular}{r@{\quad}r@{\quad}r@{\quad}r}
      \hline\hline
      $A_g$ irrep  & $B_g$ irrep & $A_u$ irrep & $B_u$ irrep  \\
       \hline
        30   &  33  &       &      \\
        42   &      &       &  38  \\
        52   &      &       &  48  \\
        61   &      &  58   &  58  \\
             &  65  &       &      \\
        68   &  69  &       &  70  \\
        73   &  73  &       &  75  \\
        78   &      &       &  76  \\
        84   &      &  81   &      \\
             &  89  &  90   &  87  \\
        93   &      &       &  93  \\
             &      &       &  94  \\
             &  99  &       &      \\
        104  &      &  104  &  106 \\ 
        108  &      &       &      \\
             &  122 &  122  &      \\     
        128  &      &       &      \\ 
        132  &      &       &  132 \\
        142  &      &       &  139 \\
             &      &       &  151 \\
             & 157  &  157  &      \\
        186  & 182  &  181  &  184 \\
             &      &       &  201 \\
        206  &  210 &  209  &      \\
             &  215 &  215  &      \\
             &  227 &  226  &      \\
        234  & 233  &  233  & 234  \\
        256  &  254 &  252  &  251 \\
        274  &      &       &  269 \\
        277  &      &       &  280 \\
        291  &      &       &      \\
             &      &       &  298 \\
        314  &      &       &      \\
             &      &       &  320 \\
      \hline \hline
    \end{tabular}
   \end{minipage}
\end{table}

\clearpage

\bibliography{BFS.bib}